\documentclass[12pt]{article}
\usepackage{amsmath,amssymb,bm,graphicx}
\usepackage{color} 
\usepackage{hyperref}
\usepackage{slashed}

\newcommand{\pslash}{\not \! p}

\newcommand{\tp}{\text{p}}

\newcommand{\sgn}{\text{sgn}}

\newcommand{\px}{{\bf p\cdot x }}
\usepackage{cite}

\begin{document}


\begin{center}
{\Large{\bf Can Oscillating Neutrino States   \\
 \vskip 0.3cm Be Formulated Universally?}}
\end{center}
\vskip .5 truecm
\begin{center}
{\bf {  Anca Tureanu}}
\end{center}

\begin{center}
\vspace*{0.4cm} 
{Department of Physics, University of Helsinki, P.O.Box 64, 
\\FIN-00014 Helsinki,
Finland
}
\end{center}
\vspace*{0.2cm} 
\begin{abstract}
A standing problem in neutrino physics is the consistent and universal definition of oscillating neutrino states
as coherent superpositions of massive neutrino states. This problem is solved in a quantum field theoretical framework of neutrino mixing developed in analogy with the Nambu--Jona-Lasinio model for the dynamical generation of nucleon masses. The massive neutrino states are Bogoliubov quasiparticles and their vacuum is a condensate of "Cooper pairs" of massless flavour neutrinos. Their superpositions as oscillating neutrino states have {\it intrinsic quantum coherence} by construction. In this quantization framework, the standard phenomenological flavour neutrino states and oscillation probability formula are validated in the ultrarelativistic approximation.
\end{abstract}

\section{Introduction}\label{intro}

The discovery of neutrino oscillations \cite{osc_exp1,osc_exp2} is the most prominent achievement of physics beyond the Standard Model. This phenomenon signals the fact that neutrinos are massive and they mix coherently, in contrast with the Standard Model massless neutrinos. Neutrino oscillations were predicted long ago \cite{Pontecorvo1,Pontecorvo2, Pontecorvo3, MNS} and their standard theoretical description has been developed in the framework of quantum mechanics \cite{Bilenky_hist}-\cite{valle}. Massive neutrinos bring about other puzzling questions, regarding their nature as Dirac or Majorana particles (see, for example, \cite{Schechter-Valle}). Plausible mechanisms of leptogenesis indicate the massive neutrinos, of either Majorana \cite{FY} or Dirac type \cite{ARS}, as responsible for the baryonic asymmetry of the Universe (for an ample review of neutrinos in cosmology, see \cite{Dolgov}). The neutrino oscillations will provide also the best test for a possible CPT violation in the leptonic sector, at DUNE and Hyper-Kamiokande \cite{AdG_CPT, GB_CPT}.

In Standard Model, neutrinos are massless and carry a $U(1)$ global quantum number called flavour. For each family of leptons (electron, muon and tau-lepton and their corresponding neutrinos), family flavour number is conserved. This conservation also implies that, for example, if an electron neutrino is produced in a process, it will always be detected as electron neutrino. The neutrinos of Standard Model are immutable due to the flavour conservation. In contrast, when we allow the neutrinos to mix and become massive, family flavour number oscillates between the production and detection, and a particle produced as muon neutrino may be detected as electron neutrino.

For simplicity and clarity of the exposition, we shall consider throughout this paper the mixing of two families of Dirac neutrinos. The extension to three families is straightforward. Majorana neutrinos can be treated as well by the procedure described below. In the standard treatment the oscillating neutrino states, customarily called "flavour states", are represented as unitary superpositions of massive neutrino states \cite{Pontecorvo3,MNS},
\begin{eqnarray}\label{states_mix}
|{\nu_e}\rangle&=&
            \cos\theta |\nu_1\rangle +\sin\theta|\nu_2\rangle,\cr
|{\nu_{\mu}}\rangle&=&        -\sin\theta|\nu_1\rangle+\cos\theta|\nu_2\rangle,
\end{eqnarray}
where $|{\nu_e}\rangle$ and $|{\nu_\mu}\rangle$ represent the electron and muon neutrino states, and $|\nu_1\rangle$ and $|\nu_2\rangle$ represent the massive neutrino states, with the masses $m_1$ and $m_2$, while $\theta$ is the mixing angle. It is also assumed that the superposition is {\it coherent}, namely that the phase difference between the two massive neutrinos is always the same (and usually taken to be zero). Having different dispersion relations, the propagating massive states develop a time-dependent phase difference, such that an electron neutrino is turned, after a macroscopic distance of propagation, into a muon neutrino. The standard oscillation probability, in the approximation that neutrinos are ultrarelativistic, is 
\begin{eqnarray}\label{prob_approx}
P_{\nu_e\nu_\mu}=&\sin^22\theta\sin^2\left(\frac{\Delta m^2}{4E}L\right), \ \ \ \Delta m^2=m_2^2-m_1^2,
\end{eqnarray}
where $E$ is the energy of the neutrinos in the beam and $L$ is the distance between the neutrino production and detection points. The details regarding the standard treatment of neutrino oscillations, including the current view on the coherence of the massive neutrino states, can be found in the monographs \cite{mohapatra}-\cite{valle}. 

The quantum mechanical formalism relies essentially on the constancy of the number of particles, therefore it cannot include the neutrino production and absorption processes. However	, all the interactions of the neutrinos are described by the Standard Model, therefore a consistent treatment of all the aspects of propagation, oscillation and interaction has to be done within a quantum field theoretical framework. How can one then connect the Lagrangian underlying the propagation and interaction of massive neutrino fields with the intuitive picture of the oscillating neutrino states presented above?

The definition of the oscillating neutrino states in quantum field theory is a standing problem ever since the neutrino oscillations were predicted by Pontecorvo \cite{Pontecorvo2}. There are several phenomenological approaches within extensions of the Standard Model  (see, e.g., \cite{GKL}--\cite{CG}  and references therein), all of them leading in the ultrarelativistic approximation to the standard oscillation formula \eqref{prob_approx}, though there exists still some debate regarding several more subtle issues in the theory of oscillation (see, for example, \cite{Akhmedov_new}).

Our understanding of the nature of oscillating neutrinos hinges on a consistent formulation of the mechanism of production/detection of {\it coherent neutrino states}. This coherence, which precludes the discrimination between different neutrino mass eigenstates, is the key element for achieving the interference leading up to the oscillation. In this paper we propose a theoretically rigorous definition of oscillating neutrino states, fulfilling the requirements of coherence and universality. 

\section{Phenomenological definitions of flavour neutrino states}

In the Standard Model, the massless neutrino fields $\psi_{\nu_l}$, with $l=e,\mu$ interact with the conservation of $U(1)$ lepton family number, for example,
\begin{eqnarray}\label{Lint}
{\cal L}_{\text{int}}=-\frac{g}{\sqrt2}\left[\bar\psi_{\nu_e}(x)\gamma_{\mu L} e(x)+\bar\psi_{\nu_\mu}(x)\gamma_{\mu L} \mu(x)\right]W^\mu+h.c.,
\end{eqnarray}
where $\gamma_{\mu L}=\gamma_{\mu}\frac{1-\gamma_5}{2}$,  $e(x)$ and $\mu(x)$ are the electron and muon fields and $W^\mu$ is the field of the $W^{\pm}$ gauge bosons. The fields $\psi_{\nu_l}$ are called for good reasons {\it flavour neutrino fields}.
The basis of the quantum field theoretical treatment of neutrino oscillations is to consider the Standard Model interaction terms and replace the fields $\psi_{\nu_l}$ by the mixed neutrino fields $\Psi_{\nu_l}$, whose quadratic Lagrangian reads:
\begin{eqnarray}\label{Lagr}
{\cal L}&=&\overline\Psi_{\nu_e}(x)i\slashed{\partial}\Psi_{\nu_e}(x) +\overline\Psi_{\nu_{\mu}}(x)i\slashed{\partial}\Psi_{\nu_{\mu}}(x)\\
&-& \left(\begin{array}{c c}
            \overline\Psi_{\nu_e}(x)&  \overline\Psi_{\nu_{\mu}}(x)
            \end{array}\right) \left(\begin{array}{c c}
            m_{ee} &m_{e\mu}\\
            m_{e\mu}&m_{\mu\mu}
            \end{array}\right)\left(\begin{array}{c}
            \Psi_{\nu_e}(x)\\
            \Psi_{\nu_{\mu}}(x)
            \end{array}\right).\nonumber
\end{eqnarray}
Upon diagonalization, \eqref{Lagr} becomes
\begin{eqnarray}\label{Lagr_diag}
{\cal L}&=&\overline\Psi_1(i\slashed{\partial}-m_1)\Psi_1 +\overline\Psi_2(i\slashed{\partial}-m_2)\Psi_2,
\end{eqnarray}
with the electron and muon neutrino fields being expressed as mixings of the massive neutrino fields as
\begin{eqnarray}\label{rotation}
\left(\begin{array}{c}
            \Psi_{\nu_e}(x)\\
            \Psi_{\nu_{\mu}}(x)
            \end{array}\right)= \left(\begin{array}{c c}
            \cos\theta &\sin\theta\\
           -\sin\theta&\cos\theta
            \end{array}\right)\left(\begin{array}{c}
            \Psi_{1}(x)\\
            \Psi_{2}(x)
            \end{array}\right),
\end{eqnarray}
where 
\begin{equation}\label{theta}
\tan2\theta=\frac{2m_{e\mu}}{m_{\mu\mu}-m_{ee}}
\end{equation}
and 
\begin{eqnarray}\label{masses}
m_1=m_{ee}\cos^2\theta+m_{\mu\mu}\sin^2\theta-m_{e\mu}\sin2\theta,\cr
m_2=m_{ee}\sin^2\theta+m_{\mu\mu}\cos^2\theta+m_{e\mu}\sin2\theta.
\end{eqnarray}
%
%
The quanta of the fields $\Psi_1$ and $\Psi_2$ represent the primary excitations of the system, i.e. massive neutrino states.

The interaction is expressed by the analogue of the Lagrangian of interaction \eqref{Lint}, in which the massless fields $\psi_{\nu_l}$ are replaced by the mixed neutrino fields $\Psi_{\nu_l}$. However, upon the diagonalization of the quadratic part by the transformation  \eqref{rotation}, the Lagrangian of interaction becomes, in terms of the massive fields:
\begin{eqnarray}\label{Lint_12}
{\cal L}_{\text{int}}=-\frac{g}{2\sqrt2}\Big[\cos\theta\bar\Psi_{1}(x)\gamma_{\mu L} e(x)+\sin\theta\bar\Psi_{2}(x)\gamma_{\mu L} e(x)\cr
-\sin\theta\bar\Psi_{1}(x)\gamma_{\mu L} \mu(x)+\cos\theta\bar\Psi_{2}(x)\gamma_{\mu L} \mu(x)\Big]W^\mu
+h.c.
\end{eqnarray}
As it stands, the Lagrangian composed of \eqref{Lint_12} and \eqref{Lagr_diag} contains only massive neutrino fields and can be easily quantized, leading to two massive neutrinos interacting both with the electron and the muon. 
Customarily, at this point the flavour neutrino states \eqref{states_mix} are introduced, though it is well known that they cannot be quanta of the flavour fields $\Psi_{\nu_l}$ (for a proof, see \cite{GKL}).

If we are to speak about neutrino oscillations, we have to be able to define the {\it coherent oscillating neutrino states}, which should be {\it associated to the fields $\Psi_{\nu_e}$ and $\Psi_{\nu_{\mu}}$}. (Such states are called in the literature "flavour neutrino states", but we shall avoid this terminology because they do not have definite family flavour number. Instead, we reserve the term of {\it flavour states} solely for the Standard Model massless neutrinos.) It is clear from the construction that these fields do not admit their own creation and annihilation operators, because they are not in definite representations of the Poincar\'e group. It is therefore necessary to develop a prescription for assigning states to the fields $\Psi_{\nu_e}$ and $\Psi_{\nu_{\mu}}$. 

One ingenious proposal in the literature has been to define the oscillating neutrino states phenomenologically, by the production or detection process in which they take part \cite{GKL,BG, G1,G2}. In this approach, it is postulated that the massive neutrinos are emitted or absorbed coherently, and the coefficients of their superposition are the matrix elements of the neutrino production/detection process. As a result, the oscillating neutrino states are process-dependent, though in the ultrarelativistic limit (which is the only limit in which neutrino oscillations have been observed~\footnote{The neutrino masses do not exceed 1 eV,  while in neutrino
oscillation experiments neutrinos with energy $E >100$ keV are detected.
}), they reduce to the "standard flavour states" of the form \eqref{states_mix}. An alternative quantum field theoretical approach is to consider the oscillating neutrinos only in intermediate states, as virtual particles \cite{GKLL, Grimus-Stock} (for reviews, see \cite{neutrino_rev,Akh-Kopp}), thus eliminating the need for defining flavour neutrino states. This solution is somewhat unnatural, in view of the macroscopic distances travelled by the oscillating neutrinos. An attempt to construct a Fock space of flavour neutrino states \cite{BV1} (see also \cite{BV2} and references therein) has been ruled out in Ref. \cite{CG}, by proving that the resulting flavour states are unphysical.

\section{Universal oscillating neutrino states}

In this paper, we propose a novel framework for the definition of oscillating neutrino states, which are universal in nature (i.e. process-independent) and coherently emitted and absorbed by definition. These states inherit a hint of the family flavour number from the Standard Model neutrino fields, which is actually their guaranty of universality. 

The quantization prescription we propose is inspired by the Bardeen--Cooper--Schrieffer (BCS) theory \cite{BCS} in Bogoliubov's treatment \cite{Bogoliubov}, or by the Nambu--Jona-Lasinio (NJL) model \cite{NJL}, with a twist due to the mixing of states. The analogy does not include the spontaneous symmetry breaking. In contrast to those models, in this case the vacuum is unique, because we start with the {\it effective theory} described by the Lagrangian \eqref{Lagr}. The procedure works irrespective of the concrete mass generation mechanism. In the most popular models, the mass terms for neutrinos are the result of the Brout--Englert--Higgs mechanism in an extension of the Standard Model. Once the vacuum is fixed for the electroweak theory with Dirac neutrino mass terms, that will be the vacuum \eqref{normalized vacuum} of this scheme and its "uniqueness" has to be understood in this sense. Thus, massive neutrinos can be viewed as Bogoliubov quasiparticles (this analogy has been earlier suggested for Majorana neutrinos in \cite{FT}, as well as for the case of neutron-antineutron oscillations \cite{FT-neutron, AT-neutron}). However, other mass generation schemes are also possible, for example the gravitationally triggered neutrino condensate \cite{GB,GB+} or the scenario in which 	the small neutrino masses emerge from a topological formulation of the gravitational anomaly  \cite{Dvali}.

\subsection{Massive neutrinos as Bogoliubov quasiparticles}

The general framework of this quantization procedure is the method of unitarily inequivalent representations, which is the basis of many fundamental results, including Haag's theorem \cite{Haag}. It has the remarkable feature that it can relate consistently the Standard Model flavour neutrino fields with the massive neutrino fields. Detailed presentations of the method can be found in Refs.  \cite{Bog-Shirk,Umezawa-book}.


The procedure is developed in the Heisenberg picture, where the time-dependent Heisenberg fields satisfy the canonical equal-time anticommutation relations as well as the equation of motion
\begin{equation}
i\partial_t\Psi(x)=[\Psi(x),H],
\end{equation}
where $H$ is the Hamiltonian of the system. The Fock space of physical, or observable, states of the Hamiltonian consists by definition of {\it free particle states}, obtained by the application of creation operators to the physical vacuum of the theory. When the Hamiltonian is expressed in terms of the creation and annihilation operators of the physical particles, it has automatically the form of a free Hamiltonian. We say then that the Heisenberg fields are realized in the Fock space of the physical free particles \cite{Umezawa-book}. This is a general feature of the Heisenberg picture. 

The scope of the method of unitarily inequivalent representations is to determine the Fock space of the free physical states of the Hamiltonian. The procedure relies on the self-consistency between the Heisenberg fields and their physical Fock space \cite{Umezawa-book} and it can be summarized as follows:

i) write down the classical Hamiltonian of the theory (corresponding to a given classical Lagrangian); 

ii) choose a set of {\it candidate free quantum fields}, based on some physical considerations, and expand the Hamiltonian in terms of their creation and annihilation operators. The Hamiltonian will usually be non-diagonal. The Fock space corresponding to the candidate fields is built on a vacuum $|0\rangle$ and its particle states describe {\it bare particles}; 

iii) diagonalize the Hamiltonian by introducing new creation and annihilation operators through Bogoliubov transformations among the operators of the initial candidate fields. The new creation and annihilation operators act on a new vacuum, which is the physical vacuum of the theory. The initial candidate fields and the fields which diagonalize the Hamiltonian are all canonical, but unitarily inequivalent (they cannot be related by a unitary transformation in the infinite volume limit).

In this quantization procedure, the states of the physical Fock space are Bogoliubov quasiparticles and
the new vacuum $|\Phi_0\rangle$ of the physical Fock space is a condensate of spinless zero-momentum pairs of bare particles (and antiparticles). Physically, the condensation arises due to some attractive interactions between the bare particles. In the case of the BCS theory, this is the interaction between the bare electrons and the phonons of the superconductor lattice, while in the NJL model it is the strong interaction between massless nucleons. For Dirac neutrinos, it is the interaction with the Higgs field. Essentially, the interaction term leads to an effective bilinear term, which is responsible for inducing the new ground state $|\Phi_0\rangle$.

In the case of neutrinos and their oscillating states, the use of this quantization method will closely parallel the NJL model. We start from the Standard Model with massless neutrino fields, but allowing also the right-handed chiral field. The massless neutrinos will be the bare particles, with their vacuum $|0\rangle$. The Yukawa interactions with the Higgs field lead effectively to the bilinear (mass) terms in \eqref{Lagr}, which break the chiral symmetry. Part of those bilinear terms break also the family lepton number symmetry. As a result, the bare vacuum $|0\rangle$ will be shifted to the physical vacuum $|\Phi_0\rangle$, which is the ground state for {\it massive neutrinos} (regarded as Bogoliubov quasiparticles, as we shall see below). All these elements will enable us finally to define coherent and universal (process-independent) oscillating neutrino states on the physical vacuum (see Sect. \ref{def_osc}).

Bellow we shall go step-by-step through the quantization procedure outlined above. 

\subsection*{i) The Hamiltonian with lepton number violation}

The Hamiltonian corresponding to the Lagrangian \eqref{Lagr} is:
\begin{eqnarray}\label{Ham}
H&=&\int d^3x \Big [-\overline\Psi_{\nu_e}(x)i\gamma^i\partial_i\Psi_{\nu_e}(x)-\overline\Psi_{\nu_{\mu}}(x)i\gamma^i\partial_i\Psi_{\nu_{\mu}}(x)\Big]\cr
&+&\int d^3x \Big [m_{ee}\overline\Psi_{\nu_e}(x)\Psi_{\nu_e}(x)+m_{\mu\mu}\overline\Psi_{\nu_\mu}(x)\Psi_{\nu_\mu}(x)\cr
&+& m_{e\mu}\left(\overline\Psi_{\nu_e}(x)\Psi_{\nu_{\mu}}(x)+\overline\Psi_{\nu_{\mu}}(x)\Psi_{\nu_e}(x)\right) \Big ]\cr
&=&H_0+H_{mass},
\end{eqnarray}
where $H_0$ is  (formally) the Hamiltonian of two massless Dirac fields and $H_{mass}$ contains the mass terms, the nondiagonal ones violating family lepton number.

\subsection*{ii) Bare fields as Standard Model massless neutrino fields} 

In the spirit of the method of unitarily inequivalent representations, we shall express the Hamiltonian in terms of the modes of the free massless fields $\psi_{\nu_e}$ and $\psi_{\nu_\mu}$, which are the Standard Model neutrino fields, i.e. solutions of the equations of motion governed by the lepton number conserving Hamiltonian $H_0$:
\begin{eqnarray}\label{Dirac_massless}
i\gamma^\mu\partial_\mu\psi_{\nu_l}(x)=0,\ \ \ \  l=e,\mu.
\end{eqnarray}
The solutions of \eqref{Dirac_massless} are written in mode expansion, at $t=0$:
\begin{eqnarray}\label{Dirac_mode_exp_l}
\psi_{\nu_l}({\bf x},0)=\int\frac{d^3p}{(2\pi)^{3/2}\sqrt{2\tp}}\sum_\lambda\Big(a_{l\lambda}({\bf p})u_\lambda({\bf p})e^{i{\bf p\cdot x }}
+b^\dagger_{l\lambda}({\bf p})v_\lambda({\bf p})e^{-i{\bf p\cdot x }}\Big),
\end{eqnarray}
where $\lambda=\pm1$ is the helicity and $\tp=|\bf p|$. The operators $a_l, a_l^\dagger, b_l, b_l^\dagger$ are creation and annihilation operators on a vacuum $|0\rangle$,
%
\begin{eqnarray}\label{naive_vac}
a_{l\lambda}({\bf p})|0\rangle=b_{l\lambda}({\bf p})|0\rangle=0, \ \ \ \ \ l=e,\mu,\end{eqnarray}
and satisfy ordinary anticommutation relations:
\begin{eqnarray}\label{ACR_ord}
\{a_{l\lambda}({\bf p}),a^\dagger_{l'\lambda'}({\bf k})\}&=&\delta_{ll'}\delta_{\lambda\lambda'}\delta({\bf p}-{\bf k}),\\
\{b_{l\lambda}({\bf p}),b^\dagger_{l'\lambda'}({\bf k})\}&=&\delta_{ll'}\delta_{\lambda\lambda'}\delta({\bf p}-{\bf k}),\nonumber
\end{eqnarray}
all the other anticommutators being zero. The states 
\begin{eqnarray}\label{bare_neutron}
a^\dagger_{e\lambda}({\bf p})|0\rangle \ \ \ \text{and }\ \ \ \ a^\dagger_{\mu\lambda}({\bf p})|0\rangle\end{eqnarray}
represent Standard Model (bare massless) electron and muon neutrinos, respectively. We assign family lepton number $+1$ to the bare neutrino states and $-1$ to the bare antineutrino states. In this sense, the Fock space of massless states built on the vacuum $|0\rangle$ is the space of {\it flavour states}. 

We proceed by going to the Schr\"odinger picture, at $t=0$, and making the identification \cite{Bogoliubov, AT-neutron,UTK,Bog-Shirk}
\begin{equation}\label{NJL}
\Psi_{\nu_l}({\bf x},0)=\psi_{\nu_l}({\bf x},0),\ \ \ l=e,\mu,
\end{equation}
in the Hamiltonian \eqref{Ham}. {\it This relation will be essential later for the definition of the oscillating neutrino states.}
The Hamiltonian \eqref{Ham}, in terms of the modes of the fields \eqref{Dirac_mode_exp_l}, reads as follows:
\begin{eqnarray}\label{Ham_modes}
H&=&\int d^3p\sum_{\lambda}\Big\{ \tp\left(a^\dagger_{e\lambda}({\bf p}) a_{e\lambda}({\bf p})+b^\dagger_{e\lambda}({\bf p})b_{e\lambda}({\bf p})+a^\dagger_{\mu\lambda}({\bf p}) a_{\mu\lambda}({\bf p})+b^\dagger_{\mu\lambda}({\bf p})b_{\mu\lambda}({\bf p})\right)\\
&+& \text{sgn}\,\lambda\, \Big[m_{ee}\left( a^\dagger_{e\lambda}({\bf p}) b^\dagger_{e\lambda}(-{\bf p})+b_{e\lambda}({\bf p})a_{e\lambda}(-{\bf p})\right)+m_{\mu\mu}\left( a^\dagger_{\mu\lambda}({\bf p}) b^\dagger_{\mu\lambda}(-{\bf p})+b_{\mu\lambda}({\bf p})a_{\mu\lambda}(-{\bf p})\right)\cr
&+&m_{e\mu}\left( a^\dagger_{e\lambda}({\bf p}) b^\dagger_{\mu\lambda}(-{\bf p})+b_{\mu\lambda}({\bf p})a_{e\lambda}(-{\bf p})+ a^\dagger_{\mu\lambda}({\bf p}) b^\dagger_{e\lambda}(-{\bf p})+b_{e\lambda}({\bf p})a_{\mu\lambda}(-{\bf p})\right)\Big]\Big\}.
\nonumber
\end{eqnarray}
%
%

\subsection*{iii) Diagonalization of the Hamiltonian and Bogoliubov transformations} 

We bring the Hamiltonian \eqref{Ham_modes} to the diagonal form
by employing a two-step procedure. First we rotate the creation and annihilation operators in \eqref{Ham_modes}:
\begin{eqnarray}\label{rotation_modes}
\left(\begin{array}{c}
            a_{e\lambda}({\bf p})\\
            a_{\mu\lambda}({\bf p})
            \end{array}\right)= \left(\begin{array}{c c}
            \cos\theta &\sin\theta\\
           -\sin\theta&\cos\theta
            \end{array}\right)\left(\begin{array}{c}
            a_{1\lambda}({\bf p})\\
            a_{2\lambda}({\bf p})
            \end{array}\right),
\end{eqnarray}
and similarly for the antineutrino operators, with the mixing angle $\theta$ given by \eqref{theta}. This is a {\it unitarily  equivalent representation of the canonical commutators} \eqref{ACR_ord}, meaning that the new operators are canonical:
\begin{eqnarray}\label{ACR_ord_new}
\{a_{i\lambda}({\bf p}),a^\dagger_{j\lambda'}({\bf k})\}&=&\delta_{ij}\delta_{\lambda\lambda'}\delta({\bf p}-{\bf k}),\\
\{b_{i\lambda}({\bf p}),b^\dagger_{j\lambda'}({\bf k})\}&=&\delta_{ij}\delta_{\lambda\lambda'}\delta({\bf p}-{\bf k}),\ \ \ i,j=1,2,\nonumber
\end{eqnarray}
and annihilate the same vacuum state $|0\rangle$.
In effect, we have introduced in this way a new set of massless Dirac fields, denoted by $\psi_{\nu_i}(x)$ with $i=1,2$, such that
\begin{eqnarray}\label{rotation_massless_fields}
\left(\begin{array}{c}
            \psi_{\nu_e}(x)\\
            \psi_{\nu_{\mu}}(x)
            \end{array}\right)= \left(\begin{array}{c c}
            \cos\theta &\sin\theta\\
           -\sin\theta&\cos\theta
            \end{array}\right)\left(\begin{array}{c}
            \psi_{1}(x)\\
            \psi_{2}(x)
            \end{array}\right),
\end{eqnarray}
which satisfy
\begin{eqnarray}\label{Dirac_massless_1,2}
i\gamma^\mu\partial_\mu\psi_{\nu_i}(x)=0,\ \ \ \  1=1,2.
\end{eqnarray}
The new massless fields $\psi_{\nu_i}$ do not have definite family lepton numbers due to the mixing \eqref{rotation_massless_fields}. In terms of the new operators $a_{i\lambda}({\bf p})$ and $b_{i\lambda}({\bf p})$, the Hamiltonian \eqref{Ham_modes} reads:
\begin{eqnarray}\label{Ham_modes_2}
H&=&\int d^3p\sum_{\lambda,i}\Big[ \tp\left(a^\dagger_{i\lambda}({\bf p}) a_{i\lambda}({\bf p})+b^\dagger_{i\lambda}({\bf p})b_{i\lambda}({\bf p})\right)\\
&+& \text{sgn}\,\lambda\, m_{i}\left( a^\dagger_{i\lambda}({\bf p}) b^\dagger_{i\lambda}(-{\bf p})+b_{i\lambda}({\bf p})a_{i\lambda}(-{\bf p})\right)\Big],
\nonumber
\end{eqnarray}
which is reminiscent of the Nambu--Jona-Lasinio Hamiltonian before diagonalization (see, for example, \cite{UTK}).

The second step is to diagonalize \eqref{Ham_modes_2} by defining the following Bogoliubov transformations:
\begin{eqnarray}\label{BT}
A_{i\lambda}({\bf p})&=&\alpha_{i\tp}a_{i\lambda}({\bf p})+\beta_{i\tp}b^\dagger_{i\lambda}(-{\bf p}),\cr
B_{i\lambda}({\bf p})&=&\alpha_{i\tp}b_{i\lambda}({\bf p})-\beta_{i\tp}a^\dagger_{i\lambda}(-{\bf p}),\ \ \ i=1,2,
\end{eqnarray}
where 
\begin{eqnarray}\label{BT_coeff}
\alpha_{i\tp}=\sqrt{\frac{1}{2}\left(1+\frac{\tp}{\Omega_{i\tp}}\right)},\ \
\beta_{i\tp}=\sgn\,\lambda\,\sqrt{\frac{1}{2}\left(1-\frac{\tp}{\Omega_{i\tp}}\right)}
\end{eqnarray}
and 
\begin{eqnarray}\label{Omega}
\Omega_{i\tp}&=&\sqrt{{\tp}^2+m_i^2}.
\end{eqnarray}
The coefficients in \eqref{BT} obey the conditions
\begin{eqnarray}\label{square_1_osc}
|\alpha_{i\tp}|^2+|\beta_{i\tp}|^2=1,\ \ \ i=1,2,
\end{eqnarray}
what insures that the transformations \eqref{BT} are canonical, such that
the new operators satisfy the canonical anticommutation relations
\begin{eqnarray}\label{ACR_new}
\{A_{i\lambda}({\bf p}),A^\dagger_{j\lambda'}({\bf k})\}&=&\delta_{ij}\delta_{\lambda\lambda'}\delta({\bf p}-{\bf k}),\cr
\{B_{i\lambda}({\bf p}),B^\dagger_{j\lambda'}({\bf k})\}&=&\delta_{ij}\delta_{\lambda\lambda'}\delta({\bf p}-{\bf k}),
\end{eqnarray}
with all the other anticommutators being zero. 

By direct calculations, performing the Bogoliubov transformation \eqref{BT} in the Hamiltonian \eqref{Ham_modes_2}, the latter is shown to aquire diagonal form in terms of the operators $A_{i\lambda}({\bf p})$ and $B_{i\lambda}({\bf p})$:
\begin{eqnarray}\label{H_osc_diag}
H&=\int d{\bf p}\sum_{\lambda,i} \Omega_{i\tp} \Big[A^\dagger_{i\lambda}({\bf p}) A_{i\lambda}({\bf p})+ B^\dagger_{i\lambda}({\bf p}) B_{i\lambda}({\bf p})\Big].
\end{eqnarray}
This is the Hamiltonian of two free Dirac fields of definite masses $m_1$ and $m_2$, and it is in accord with the diagonal expression of the Lagrangian, eq. \eqref{Lagr_diag}. 

The operators introduced in \eqref{BT} define a new vacuum state, $|\Phi_0\rangle$,
\begin{equation}\label{AB_vacuum}
A_{i\lambda}({\bf p})|\Phi_0\rangle =B_{i\lambda}({\bf p})|\Phi_0\rangle =0,\ \ \ i=1,2,
\end{equation}
which represents the physical vacuum of the theory. The physical neutrino states are Bogoliubov quasiparticles, of Dirac type, with the definite masses $m_1$ and $m_2$ given by \eqref{masses}. In this way, we have found the mode expansion of the free massive neutrino fields $\Psi_i({\bf x},0),\ i=1,2$. The evolution of these Heisenberg fields is given by
\begin{equation}
e^{iHt}\Psi_i({\bf x},0)e^{-iHt}=\Psi_i({\bf x},t), \ \ \ i=1,2,
\end{equation}
with $H$ in the form \eqref{H_osc_diag}.
The corresponding creation and annihilation operators evolve as
\begin{eqnarray}
A_i({\bf p},t)&=&e^{iHt}A_i({\bf p})e^{-iHt}=A_i({\bf p})e^{-i\Omega_{i\tp}t},\cr
A_i^\dagger({\bf p},t)&=&e^{iHt}A_i^\dagger({\bf p}) e^{-iHt}=A_i^\dagger({\bf p})e^{i\Omega_{i\tp}t},
\end{eqnarray}
and similarly for $B_i({\bf p},t)$ and $B_i({\bf p},t)^\dagger$.
Thus, the physical time-dependent massive Dirac neutrino fields will be expressed as:
\begin{eqnarray}\label{Majorana_mode_exp_H}
\Psi_i({\bf x},t)=\int\frac{d^3p}{(2\pi)^{\frac{3}{2}}\sqrt{2\Omega_{i\tp}}}\sum_{\lambda}\Big(A_{i\lambda}({\bf p})U_{i\lambda}({\bf p})e^{-i(\Omega_{i\tp} t-{\bf p\cdot x })}+B^\dagger_{i\lambda}({\bf p})V_{i\lambda}({\bf p})e^{i(\Omega_{i\tp} t-{\bf p\cdot x })}\Big),\nonumber\\
\end{eqnarray}
with the spinors $U_{i\lambda}({\bf p}), V_{i\lambda}({\bf p})$ satisfying the equations 
\begin{eqnarray*}(\pslash-m_i)U_{i\lambda}({\bf p})=0,\\
(\pslash+m_i)V_{i\lambda}({\bf p})=0.
\end{eqnarray*}
Consequently, the fields $\Psi_i({\bf x},t)$ satisfy the free massive Dirac equations,
\begin{eqnarray}\label{Dirac_massive}
(i\gamma^\mu\partial_\mu-m_i)\Psi_{i}(x)=0,\ \ \ \  i=1,2.
\end{eqnarray}

The physical vacuum $|\Phi_0\rangle$ is not annihilated by the operators $a_i({\bf p}), b_i({\bf p})$, nor by $a_l({\bf p}), b_l({\bf p})$, therefore it is a different state from $|0\rangle$. Using \eqref{BT}, \eqref{BT_coeff}, \eqref{AB_vacuum} and \eqref{square_1_osc}, we find (see, for example, \cite{NJL,AT-neutron}) that $|\Phi_0\rangle$ is a coherent superposition, or a condensate of "Cooper pairs" of massless neutrino states:
\begin{eqnarray}\label{normalized vacuum}
|\Phi_0\rangle =\ \Pi_{i,{\bf p},\lambda}\  \left(\alpha_{i\tp} -\beta_{i\tp}\,a_{i\lambda}^\dagger({\bf p})b_{i\lambda}^\dagger(-{\bf p})\right)|0\rangle.
\end{eqnarray}
Let us calculate the inner product of the two vacua, using  \eqref{normalized vacuum} and taking into account \eqref{naive_vac} and \eqref{BT_coeff}: 
\begin{eqnarray}
&&\langle 0|\Phi_0\rangle = \Pi_{i,{\bf p},\lambda}\ \alpha_{i\tp}=\Pi_{i,{\bf p},\lambda}\ \left(1+\frac{\tp}{\Omega_{i\tp}}\right)^{1/2},
\end{eqnarray}
which vanishes as $\exp\left[-(m_1^2+m_2^2)\int d{ \tp}\right]$, in the infinite momentum limit. As a result, the two vacua are orthogonal:
\begin{equation}
\langle 0|\Phi_0\rangle =0.
\end{equation}
Consequently, the Fock spaces built on the vacua $|0\rangle$ and $|\Psi_0\rangle$ do not contain any common states. The physical Fock space is the one containing the vacuum condensate $|\Psi_0\rangle$, and its elements are massive neutrino states. It should be emphasized that all the operators of type $a_l,b_l$, as well as $a_i, b_i$, when acting on the vacuum $|\Psi_0\rangle$, create massive particles. This can be seen by inverting the Bogoliubov transformations \eqref{BT}. (In contrast, the operators of the type $A_i,B_i$ create massless particles when acting on the vacuum $|0\rangle$.)

Let us emphasize that, in spite of the fact that we started with the Hamiltonian written in terms of the "flavour neutrino fields" $\Psi_{\nu_l}$ and made the Schr\"odinger picture identification \eqref{NJL}, finally we ended up with the massive neutrino fields $\Psi_i$ as physical fields. This shows once more that it is impossible to identify proper creation and annihilation operators, or states, of the fields  $\Psi_{\nu_l}$.


\subsection{Oscillating neutrino states and their transition probability}\label{def_osc}

Although the fields $\Psi_{\nu_e}$ and $\Psi_{\nu_\mu}$ do not have their own Fock spaces, we can define a prescription for {\it associating} to these fields properly defined states, which will be called below {\it oscillating neutrino states}. The rule of association has to satisfy several indispensable conditions:

\begin{enumerate}

\item the oscillating states have to be defined on the physical vacuum $|\Psi_0\rangle$, namely the vacuum of the massive neutrinos;

\item in the limit when the family lepton-number violating interaction vanishes (i.e. $m_1,m_2\to0$), one recovers the massless flavour neutrino states defined on the vacuum $|0\rangle$.
\end{enumerate}

The prescription we propose is to generalize the {\it bona fide} flavour neutrino states, defined on the bare vacuum $|0\rangle$, as $a_{l\lambda}^\dagger({\bf p})|0\rangle$. The operators $a_{l\lambda}^\dagger({\bf p})$ carry a definite family flavour number, and for this reason we shall adopt them as "neutrino creation operators" also on the physical vacuum $|\Phi_0\rangle$. In view of the relation 
\begin{equation}\label{bare_neutrino_creation}
\frac{1}{\sqrt{2\tp}}\left(\int\frac{d^3x}{(2\pi)^{3/2}}e^{i\px}\bar\psi_{\nu_l}({\bf x},0) \right)\gamma_0u_\lambda({\bf p})=a_{l\lambda}^\dagger({\bf p})
\end{equation}
and the Schr\"odinger picture identification \eqref{NJL},
$\Psi_{\nu_l}({\bf x},0)=\psi_{\nu_l}({\bf x},0),\ l=e,\mu,$
we define in an universal manner the oscillating neutrino states associated with the fields $\Psi_{\nu_l}$:
\begin{eqnarray}\label{nu_l_state}
|\nu_l({\bf p},\lambda)\rangle&\equiv&a_{l\lambda}^\dagger({\bf p})|\Phi_0\rangle\\
&=& \frac{1}{\sqrt{2\tp}}\left(\int\frac{d^3x}{(2\pi)^{3/2}}e^{i\px}\bar\Psi_{\nu_l}({\bf x},0) \right)\gamma_0u_\lambda({\bf p})|\Phi_0\rangle.\nonumber
\end{eqnarray}
This definition satisfies the two consistency requirements stated above.
Using \eqref{nu_l_state} together with \eqref{rotation_modes}, \eqref{AB_vacuum} and the inverses of the Bogoliubov transformations \eqref{BT}, we find the expressions for the {\it oscillating electron and muon neutrino states as coherent superpositions of the massive neutrino states with equal momenta}, $|\nu_{1\lambda}({\bf p})\rangle$ and $|\nu_{2\lambda}({\bf p})\rangle$:
\begin{eqnarray}\label{nu_e_state}
|\nu_e({\bf p},\lambda)\rangle&=&\left(\cos\theta\alpha_{1\tp}A^\dagger_{1\lambda}({\bf p})+\sin\theta\alpha_{2\tp}A^\dagger_{2\lambda}({\bf p})\right)|\Phi_0\rangle,\cr
&=&\cos\theta\alpha_{1\tp}|\nu_{1\lambda}({\bf p})\rangle+\sin\theta\alpha_{2\tp}|\nu_{2\lambda}({\bf p})\rangle
\end{eqnarray}
and
\begin{eqnarray}\label{nu_mu_state}
|\nu_\mu({\bf p},\lambda)\rangle&=&\left(-\sin\theta\alpha_{1\tp}A^\dagger_{1\lambda}({\bf p})+\cos\theta\alpha_{2\tp}A^\dagger_{2\lambda}({\bf p})\right)|\Phi_0\rangle\cr
&=&-\sin\theta\alpha_{1\tp}|\nu_{1\lambda}({\bf p})\rangle+\cos\theta\alpha_{2\tp}|\nu_{2\lambda}({\bf p})\rangle,
\end{eqnarray}
with the coefficients given by \eqref{BT_coeff}--\eqref{Omega}. We emphasize that the oscillating neutrino states are not orthogonal to each other, unlike the standard states \eqref{states_mix}. A priori, there is no principle to enforce the orthogonality of the  electron and muon neutrino states. The only quantum field theoretical requirement is that the massive neutrino states be orthogonal, and they are. 

The generalization to three neutrino mixing is straightforward: the flavour neutrino fields in terms of the massive fields $\Psi_{i}$, $i=1,2,3$ are written as
$$
\Psi_{\nu_l}=\sum_{i=1,2,3} U_{li}\Psi_{i},
$$
where $U_{li}$ are the elements of the unitary Pontecorvo--Maki--Nakagawa--Sakata mixing matrix. Using exactly the same procedure as above, we find the oscillating neutrino states in terms of the massive states of $\Psi_{i}$:
\begin{eqnarray}\label{3-nu_state}
|\nu_l({\bf p},\lambda)\rangle=\sum_{i=1,2,3} U^*_{li}\alpha_{i\tp}|\nu_{i\lambda}({\bf p})\rangle, \ \ \ l=e,\mu,\tau,
\end{eqnarray}
where the coefficients $\alpha_{i\tp}$ have the same expressions as in \eqref{BT_coeff}, but this time there are three of them, with $\Omega_{i\tp}$ given by the corresponding masses $m_i$, $i=1,2,3$.
We took advantage of the fact that, in spite of the $a$ and $b$-type operators not being creation and annihilation operators on the physical vacuum $|\Phi_0\rangle $, they do act on this vacuum, with the action defined through the inverse Bogoliubov transformations. 

Thus, oscillating neutrino states, associated with the fields $\Psi_{\nu_l}$ involved in the weak interactions, are naturally defined on the physical Fock space of massive neutrinos. 
By construction, they have inbuilt coherence, ensured by the coherence of the vacuum condensate \eqref{normalized vacuum}. The interaction leading to the effective mass terms in \eqref{Lagr} dresses the 
bare flavour neutrinos and transforms them into physical states. In this process of clothing \cite{Bog-Shirk}, effectively encoded in the vacuum condensate, the neutrinos gain mass and lose family flavour number.

It should be pointed out that the equations \eqref{3-nu_state} do not represent a "change of basis", from the "massive states basis" to the "oscillating states basis". The oscillating states do not exist independently of the massive states. Only the latter form a basis. As a result, inverting \eqref{3-nu_state} in order to express a massive state as a "superposition of oscillating states" is not a justifiable operation. 

The oscillation amplitude between the two types of neutrinos is obtained by letting the electron neutrino state evolve and sampling the amount of muon neutrino in it at an arbitrary time $t$:
\begin{equation}\label{trans_1}
{\cal A}_{\nu_e\to\nu_\mu}=\langle  \nu_\mu({\bf p},\lambda)|\nu_e({\bf p},\lambda),t\rangle\equiv\langle \nu_\mu({\bf p},\lambda)| e^{-iHt}|\nu_e({\bf p},\lambda)\rangle.
\end{equation}
Using \eqref{nu_e_state} and \eqref{nu_mu_state}, 
as well as the Hamiltonian in the form \eqref{H_osc_diag} and its action on the massive neutrino states $H\,A_{i\lambda}^\dagger({\bf p})|\Phi_0\rangle =\Omega_{i\tp}\ A_{i\lambda}^\dagger({\bf p})|\Phi_0\rangle$, we obtain:
\begin{eqnarray}\label{trans_2}
{\cal A}_{\nu_e\to\nu_\mu}=\frac{1}{2}\sin2\theta\Big[-(\alpha_{1\tp})^2e^{-i\Omega_{1\tp}t}+(\alpha_{2\tp})^2e^{-i\Omega_{2\tp}t}\Big],
\end{eqnarray}
with the various coefficients and energies given by \eqref{BT_coeff}--\eqref{Omega}. This is the general expression, valid for any values of the particle momenta and mass parameters in the Lagrangian \eqref{Lagr}.
To come to a more familiar expresion of the transition amplitude, we expand \eqref{trans_2} to the second order in $m_i/\tp$. In this order,
\begin{eqnarray}\label{trans_approx}
{\cal A}_{\nu_e\to\nu_\mu}=\frac{1}{2}\sin 2\theta e^{-i\tp t}\Big[-\left(1-\frac{1}{4}\frac{m_1^2}{\tp^2}\right)^2e^{-i\frac{m_1^2}{2\tp}t}
+\left(1-\frac{1}{4}\frac{m_2^2}{\tp^2}\right)^2e^{-i\frac{m_2^2}{2\tp}t}\Big].
\end{eqnarray}
We note that the transition amplitude ${\cal A}_{\nu_e\to\nu_\mu}$ is never zero, i.e. there is always a small portion of muon neutrino in the electron neutrino and vice-versa. This is due to the fact that the oscillating neutrino states are not orthogonal, but the departure from orthogonality is extremely tiny, of the order of $m_i^2/p^2$. As a result, the probability of an  electron neutrino to interact with muons, for example, is theoretically nonvanishing but experimentally inobservable.

In the ultrarelativistic approximation, we discard the terms of second order in $m_i/\tp$ and obtain
\begin{eqnarray}\label{trans_standard}
{\cal A}_{\nu_e\to\nu_\mu}=\frac{1}{2}\sin 2\theta e^{-i\tp t}\Big[-e^{-i\frac{m_1^2}{2\tp}t}+e^{-i\frac{m_2^2}{2\tp}t}\Big],
\end{eqnarray}
recovering in this limit the standard oscillation probability,
\begin{eqnarray}\label{prob_approx_2}
P_{\nu_e\to\nu_\mu}=&\sin^22\theta\sin^2\left(\frac{\Delta m^2}{4\tp}t\right), \ \ \ \Delta m^2=m_2^2-m_1^2.
\end{eqnarray}

\section{Outlook}
The universal definition of the oscillating neutrino states is the first step towards formulating the quantum field theoretical mechanism of their coherent production and absorption. We propose for the first time a prescription for constructing  {\it intrinsically coherent neutrino states} \eqref{nu_l_state}, by establishing a one-to-one correspondence with the Standard Model massless neutrino states. This construction can be implemented for the mixing of any number of Dirac or Majorana neutrinos and it represents a novel conceptual framework for the description of neutrino oscillations. The present formalism will lead to a deeper understanding of coherence and decoherence of oscillating particles, as well as oscillations of nonrelativistic neutrinos. At the same time, the elucidation of the unitarity and relativistic covariance of the mechanism of interaction of oscillating states is of utmost importance. 
%
%
%

There is a commonly held belief that the Lagrangian and Hamiltonian formalism lead to the same results. However, it should not be forgotten that when it comes to quantization, the Hamiltonian formalism is the repository of the first principles, to which we have to return whenever we analyse an unconventional physical situation, like particle oscillations. In general, the Hamiltonian formalism, in particular the method of unitarily inequivalent representations, brings out a richer dynamical picture than the simple diagonalization of the Lagrangian \cite{Bog-Shirk}. The theoretical framework proposed here is no exception: by Lagrangian diagonalization, we cannot go further than the eqs. \eqref{Lagr_diag} and \eqref{Lint_12}, which show  massive neutrino fields in interaction with charged lepton fields, with the flavour neutrino fields completely obliterated. In contrast, the Hamiltonian quantization formalism developed in this paper gives a prominent role to the flavour neutrino fields $\Psi_{\nu_l}$, $l=e,\mu$, by the eq. \eqref{NJL}. This makes possible to associate to them the oscillating neutrino states \eqref{nu_l_state}, which display intrinsic quantum coherence as superpositions of massive neutrino states.

The oscillating neutrino states thus defined \eqref{nu_l_state} differ from the standard neutrino states \eqref{states_mix}. Nevertheless, the present quantization prescription does not contradict the standard phenomenological treatment, since the corrections to the vacuum oscillations are negligible in the ultrarelativistic approximation, in which all the oscillation experiments have been performed. Thus, the standard formalism is validated as the limit of a conceptually more rigorous framework. It is interesting to note that the coherent states introduced by Klauder \cite{Klauder}, Sudarshan \cite{Sudarshan} and Glauber \cite{Glauber} and used in quantum optics are also not mutually orthogonal (see also Ref. \cite{IZ} for a presentation of the coherent state formalism in the particle physics context). The lack of orthogonality appears to be the price for coherence. 

The quantitative differences between the oscillating neutrino states \eqref{nu_l_state} and the standard flavour states \eqref{states_mix} are more pronounced for nonrelativistic neutrinos. This may suggest possible effects for the planned PTOLEMY experiment \cite{Ptolemy} for the detection of the cosmic neutrino background. However, since the relic neutrinos are considered to have decohered and to propagate as massive states without oscillating, their behaviour in interaction is not affected by the present formalism. It would be interesting to find testable situations in which nonrelativistic neutrinos do oscillate.


Moreover, it is quite plausible that this approach may lead to enhanced corrections to the MSW effect \cite{W,MS}. The quantitative differences could be even more significant for neutrinos in extreme conditions, like supernova neutrinos \cite{SN_neutrinos}. We shall consider these aspects in a future communication.

\subsection*{Acknowledgments}

I am grateful to Gabriela Barenboim, Samoil Bilenky, Masud Chaichian, Aleksandr Dolgov, Kazuo Fujikawa and Aleksandr Studenikin for inspiring discussions and valuable comments.

\end{document}